\documentclass{cernrep}
\usepackage[utf8x]{inputenc}
\usepackage{hyperref}
\newcommand{\xpom}{{x_\mathbb{P}}}
\newcommand{\der}{\mathrm{d}}

\begin{document}
\title{Photoproduction prospects at the EIC}
\author{Heikki Mäntysaari}
\institute{Physics Department, Brookhaven National Laboratory, Upton, NY 11973, USA}

\begin{abstract}
We discuss how photoproduction and electroproduction processes will provide a precise tool to study nuclear structure in the small-$x$ region in the future Electron-Ion Collider. In particular, we emphasize how exclusive vector meson production at the EIC can be used to accurately map the gluonic density profile of nuclei, including its event-by-event fluctuations. In addition, we show how measurements of the total diffractive cross section will be sensitive to saturation effects.
\end{abstract}


\maketitle

\section{Introduction}
The proposed Electron-Ion Collider (EIC) in the United States will be the first collider ever to study the inner structure of both protons and nuclei at high energy~\cite{Accardi:2012qut}. 
At the EIC, the electron beam probes protons and nuclei in deep inelastic scattering (DIS) processes. These events experimentally and theoretically cleaner compared to for example proton-nucleus collisions, making it possible to study the physics of strong interactions in unprecedented accuracy. 

Some key parts of the wide physics program of the EIC are laid out in the White paper~\cite{Accardi:2012qut} and in the EIC science case~\cite{Boer:2011fh} review. Recently, as the machine design evolves, the center-of-mass energy required by some of the most important measurements has been studied in detail~\cite{Aschenauer:2017jsk}. In this talk, the possibility to probe the saturated gluonic matter at small Bjorken-$x$ in diffractive events is discussed. 

HERA measurements of the proton structure~\cite{Abramowicz:2015mha} resulted in an accurate picture of the proton in terms of its partonic content as a function of the longitudinal momentum fraction carried by the quarks and gluons. One striking observation has been the rapid rise of the gluon distribution towards small $x$. This raise would eventually violate unitarity, and non-linear effects must  limit the growth of the gluon density at very small $x$. These non-linear \emph{saturation} phenomena are included in the Color Glass Condensate (CGC) effective theory of QCD valid in the small-$x$ region~\cite{Gelis:2010nm}. Observing signatures of the saturated dense gluonic matter, described by CGC, is one of the key tasks for the EIC. 

In diffractive processes a system of particles (or just e.g. a vector meson) is produced such that there is a large rapidity gap between the system and the beam remnants. In perturbative QCD at leading order, this requires two gluons to be exchanged with the target, as there can not be an exchange of net color charge. Thus, the cross section for such a  process is proportional to the target gluon density \emph{squared}, making diffractive scattering a sensitive probe of the gluonic structure of the nucleus. In addition, in exclusive processes where only one particle is produced, it is possible to study the spatial distribution of quarks and gluons in the proton and nuclei as the transverse momentum is Fourier conjugate to the impact parameter.

\section{Accessing saturation region at the EIC}
\begin{figure}[tb]
    \centering
    \begin{minipage}[b]{.45\textwidth}
        \centering
       \includegraphics[width=6cm]{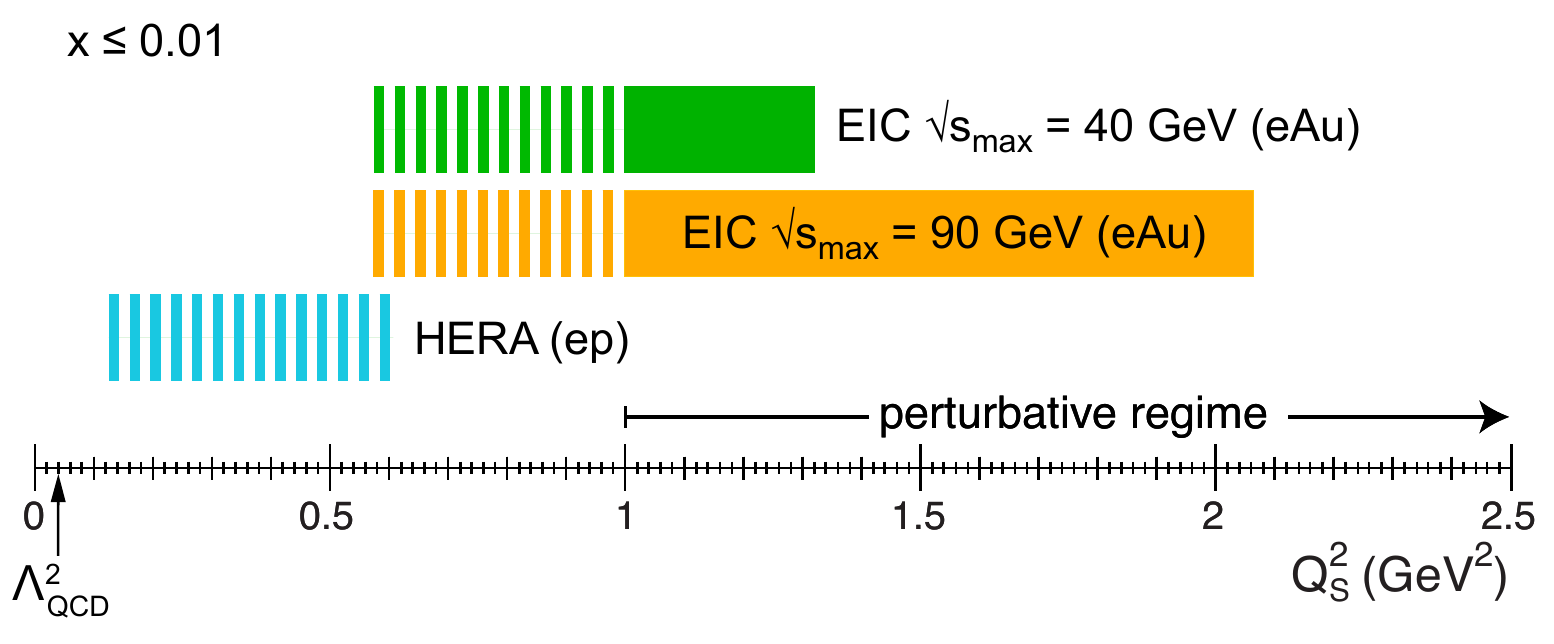}
\caption{Saturation scales $Q_s^2$ reached at the EIC in electron-nucleus collisions, compared to the ones accessed at HERA in electron-proton scattering. Figure from Ref.~\cite{Aschenauer:2017jsk}.}
\label{fig:qs_hera_eic}
    \end{minipage} 
    \hfil
    \begin{minipage}[b]{0.45\textwidth}
        \centering
	     \includegraphics[width=0.9\textwidth]{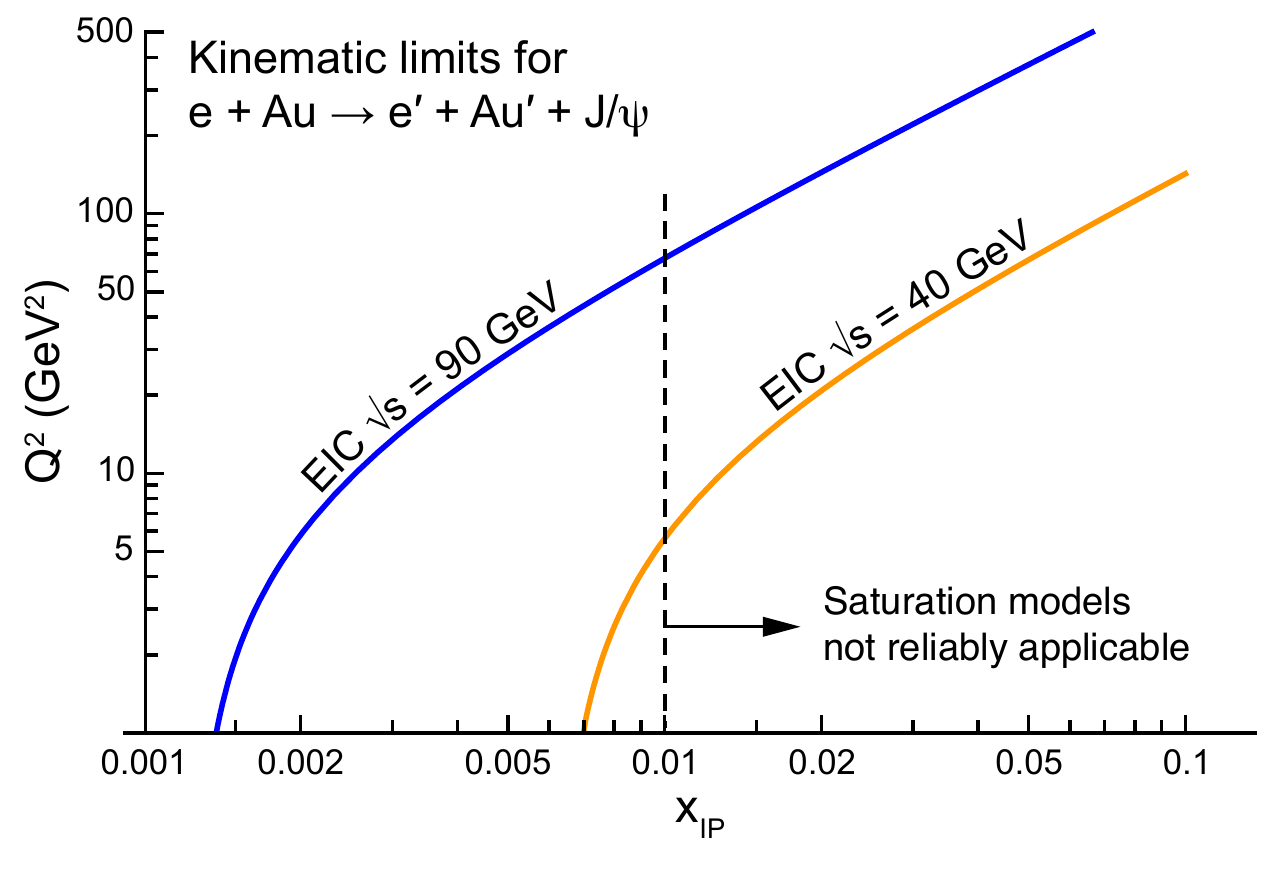} 
				\caption{Kinematical coverage for the exclusive $J/\Psi$ production at the EIC. Figure from Ref.~\cite{Aschenauer:2017jsk}.}
		\label{fig:jpsi_coverage}
	\vspace{\baselineskip}%
    \end{minipage}
\end{figure}

The Electron-Ion Collider will collide electrons with protons and a variety of different nuclei. In the electron-proton scattering, the proposed maximum center-of-mass energies are $\sqrt{s}=141$ GeV, which is the region already covered by HERA. The question whether HERA saw saturation or not is not completely settled. However, the scale at which the non-linear saturation phenomena become important, $Q_s^2$ (at given $x$), scales like $A^{1/3}$ as there are $\sim A^{1/3}$ overlapping nucleons. Thus, even at smaller photon-nucleon center of mass energies than in electron-proton mode, it is possible to access the region where saturation scale is large and in the perturbative region by using heavy nuclei. In the electron-nucleus mode, the proposed maximum energy for the EIC is $\sqrt{s}=40\dots 90$ GeV. The saturation scales accessed in electron-gold collisions at the EIC, compared to the HERA electron-proton scattering, are illustrated in Fig.~\ref{fig:qs_hera_eic}. As saturation scales probed at the EIC can be up to $2 $ GeV$^2$, it is expected that the saturation effects are clearly visible.

\section{Diffractive scattering}

In diffractive sattering process the incoming lepton emits a virtual photon (virtuality $Q^2$) which interacts with the target hadron (momentum $P$) and forms 
the final state system $X$ after gaining longitudinal momentum $\xpom P$ from the target. Here $\mathbb{P}$ stands for the interpretation that a color neutral object ``\emph{pomeron}'' is exchanged between the photon and the target. 
In exclusive processes, the system $X$ is just a single particle, usually a vector meson like $J/\Psi$ or $\rho$. The advantage of $J/\Psi$ is that its large mass allows perturbative treatment also at low $Q^2$, but it is not too heavy to lie outside the kinematical reach of an EIC even at small $x$, as demonstrated in Fig.~\ref{fig:jpsi_coverage}.

\begin{figure}[tb]
    \centering 
    \begin{minipage}[b]{0.45\textwidth}
        \centering

   \includegraphics[width=0.9\textwidth]{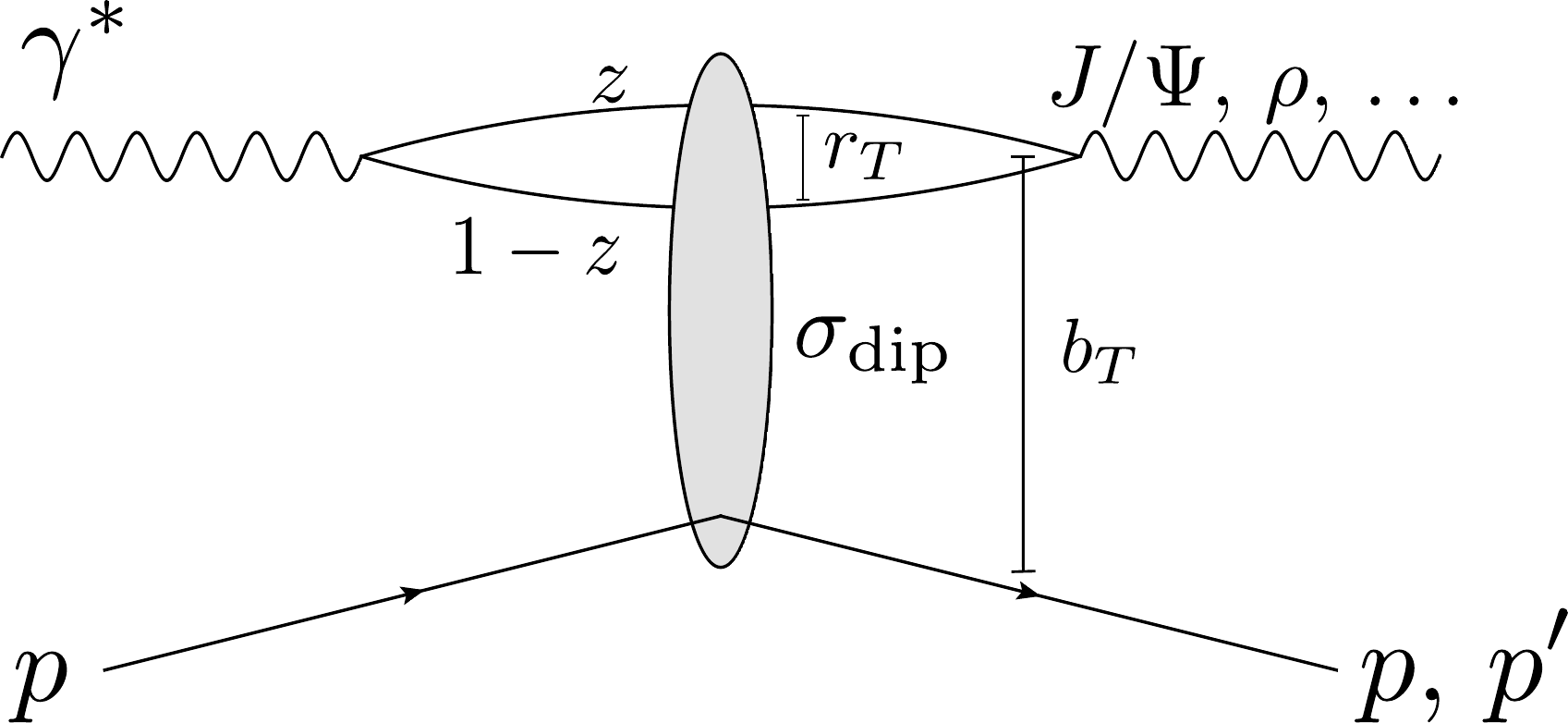} 
				\caption{Diffractive scattering in dipole picture.}
		\label{fig:dipolepicture}

	\vspace{\baselineskip}%
    \end{minipage}
    \hfill
     \begin{minipage}[b]{0.45\textwidth}
        \centering
        \includegraphics[width=0.9\textwidth]{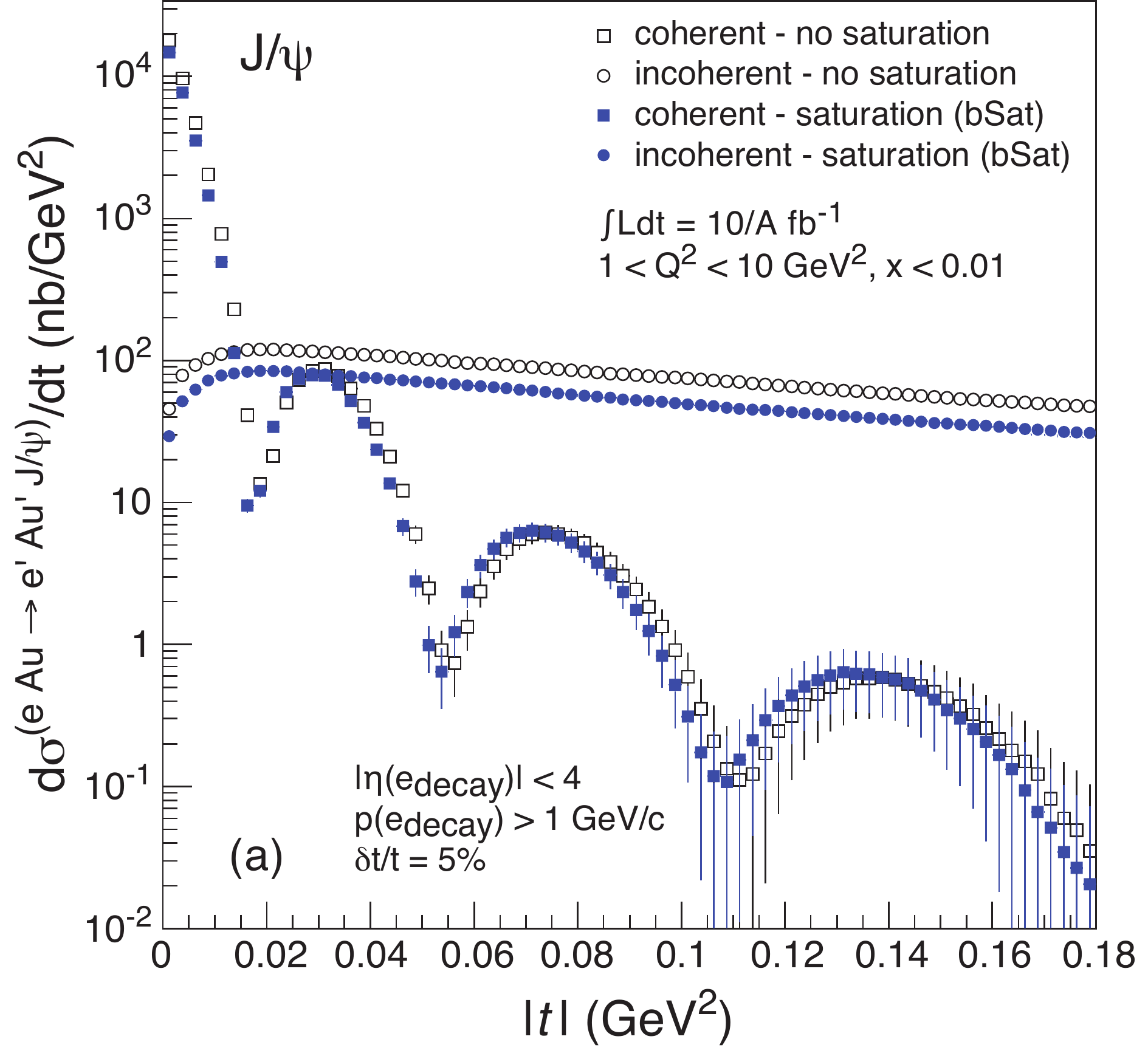}
        \caption{Simulated coherent and incoherent diffractive $J/\Psi$ production at the EIC from Ref.~\cite{Toll:2012mb}.}
        \label{fig:jpsi_eic_dt}
		
    \end{minipage}
    
\end{figure}

A convenient way to describe the scattering process is to look at the scattering in the frame where the photon is very energetic, in the so called dipole picture. In this frame, the scattering process goes like in Fig.~\ref{fig:dipolepicture}. First, the virtual photon fluctuates into a quark-antiquark dipole long before the scattering takes place. This $\gamma^* \to q\bar q$ splitting is described by the virtual photon wave function $\Psi$. The produced dipole then scatters elastically off the target (dipole scattering given by the dipole-target cross section $\sigma_\text{dip}$), and forms the final state $X$. In case of vector meson production, the $q\bar q \to $ vector meson transition is described by the vector meson wave function $\Psi_V$. In case of inclusive diffraction, where the final state is specified by the mass of the system $M_X^2$, one has to also take into account $q\bar q g$ Fock states of the virtual photon wave function~\cite{Kowalski:2008sa}.

Let us first study exclusive vector meson production (for exclusive photon production, or Deeply Virtual Compton Scattering, see Ref.~\cite{Aschenauer:2013hhw}). These events can be divided into two categories.  The total diffractive cross section is  proportional to the squared diffractive scattering amplitude $A$, averaged over all the possible target configurations $\langle |A|^2\rangle$. On the other hand, if we require that the target remains in the same quantum state despite the scattering, the cross section becomes (see Refs.~\cite{Good:1960ba,Miettinen:1978jb})
\begin{equation}
\frac{\der \sigma^{\gamma^* A \to VA}}{\der t} = \frac{1}{16\pi} \left | \langle A \rangle \right |^2.
\end{equation}
Subtracting this contribution from the total diffractive cross section we obtain the cross section for the process in which the target is left in a different quantum state after the scattering, meaning that it breaks up. This is called incoherent diffraction, and the cross section  is
\begin{equation}
\frac{\der \sigma^{\gamma^* A \to VA}}{\der t} = \frac{1}{16\pi} \left( \langle |A|^2\rangle -  \left | \langle A \rangle \right |^2 \right).
\end{equation}
Because the incoherent cross section is the variance of the scattering amplitude, it measures the amount of fluctuations in the target wave function in the impact parameter space (see e.g.  Refs~\cite{Frankfurt:1993qi,Caldwell:2009ke,Mantysaari:2016ykx} and references therein).

The diffractive scattering amplitude in the dipole picture can be written as~\cite{Kowalski:2006hc}
\begin{multline}
\label{eq:diffractive_amp}
A^{\gamma^* p \to V p}_{T,L}(\xpom,Q^2, \Delta) = i\int \der^2 r_T \int \der^2 b_T \int \frac{\der z}{4\pi}  
  (\Psi^*\Psi_V)_{T,L}(Q^2, r_T,z) \\
  \times 
  e^{-i[b_T - (1-z)r_T]\cdot \Delta}  \frac{\der \sigma_{\text{dip}}(r_T,b_T,\xpom)}{\der^2 b_T}.
\end{multline}
Here $\Delta$ is the transverse momentum of the produced vector meson. The overlap between the virtual photon and the vector meson wave functions is $\Psi^*\Psi_V$, given the virtual photon virtuality $Q^2$, dipole size $r_T$ and the longitudinal momentum fraction of the photon carried by the quark $z$.

The EIC capabilities to measure exclusive vector meson production were simulated in Ref.~\cite{Toll:2012mb}, and the simulation results for $J/\Psi$ production are shown in Fig.~\ref{fig:jpsi_eic_dt} as a function of $-t \approx \Delta^2$. Due to the large mass of the $J/\Psi$, small dipole sizes dominate the process and the saturation effects are not especially large (but somewhat enhanced in the incoherent process, see also Ref.~\cite{Lappi:2010dd}) unlike in case of lighter meson production. As the coherent cross section is obtained by Fourier transforming the dipole amplitude from coordinate space to momentum space (see Eq.~\eqref{eq:diffractive_amp}), Fourier transforming the measured cross section back to coordinate space gives the nuclear density profile in the coordinate space. The projected accuracy of the EIC is shown to be  good enough to reconstruct the nuclear density profile accurately, as demonstrated in Ref.~\cite{Toll:2012mb}. From the incoherent cross section one can extract the amount of fluctuations and study at which distance scales these fluctuations take place, as demonstrated in Refs.~\cite{Mantysaari:2016ykx,Mantysaari:2017dwh}.

If instead of $J/\Psi$ one studies the production of lighter mesons like $\rho$ or $\phi$, then the dominant contribution comes from larger dipoles that make these processes more sensitive to saturation effects. However, due to the small mass, there is no large scale which makes perturbative calculations unreliable at low $Q^2$, and the EIC capabilities to get to higher $Q^2$ are necessary to fully take the advantage of the possibility to measure exclusive production of different particle species.

In inclusive diffraction, one can study the scattering process as a function of the invariant mass of the produced system $M_X^2$. The advantage is that this makes it possible to probe both $q\bar q$ dipole-nucleus scattering (low $M_X^2)$ and $q\bar q g $ dipole-nucleus scattering (large $M_X^2$)~\cite{Kowalski:2008sa}.
The saturation picture calculations have a characteristic prediction for the nuclear effects in inclusive diffraction. First, the total cross section for the production of small invariant mass systems is enhanced in the nucleus relative to a proton. On the other hand, because the strong color field of the nucleus absorbs $q\bar q g$ dipole more strongly compared to the proton, the diffractive cross section at large $M_X$ is suppressed in the nucleus relative to proton~\cite{Kowalski:2008sa}. This is demonstrated in Fig.~\ref{fig:inclusive_diffraction}, where the nuclear modification to the diffractive structure function $F^D$ is shown (which is proportional to the total diffractive cross section). The suppression of $q\bar q g$ dipoles at large invariant masses is expected to be visible at the EIC energies~\cite{Aschenauer:2017jsk}.

\section{Accessing rare configurations}

\begin{figure}[tb]
    \centering 
    
    \begin{minipage}[b]{0.45\textwidth}
        \centering
          \includegraphics[width=0.9\textwidth]{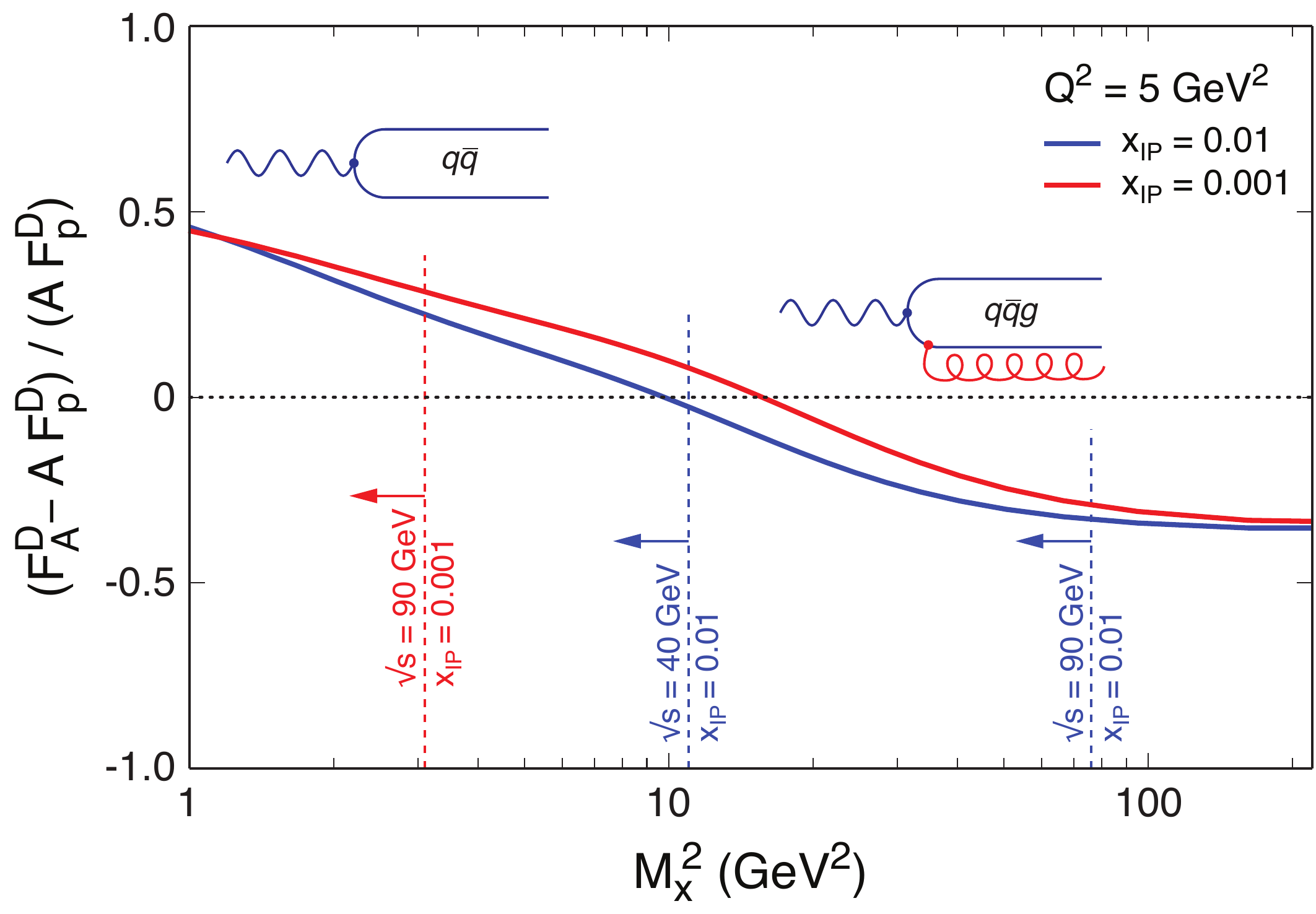} 
				\caption{Nuclear modification to the diffractive structure function as a function of the produced system mass. Dashed lines show kinematical limits for  the two different EIC energies. Figure from Ref.~\cite{Aschenauer:2017jsk}.}
		\label{fig:inclusive_diffraction}

    \end{minipage}
    \hfill
     \begin{minipage}[b]{0.45\textwidth}
        \centering
  \includegraphics[width=0.9\textwidth]{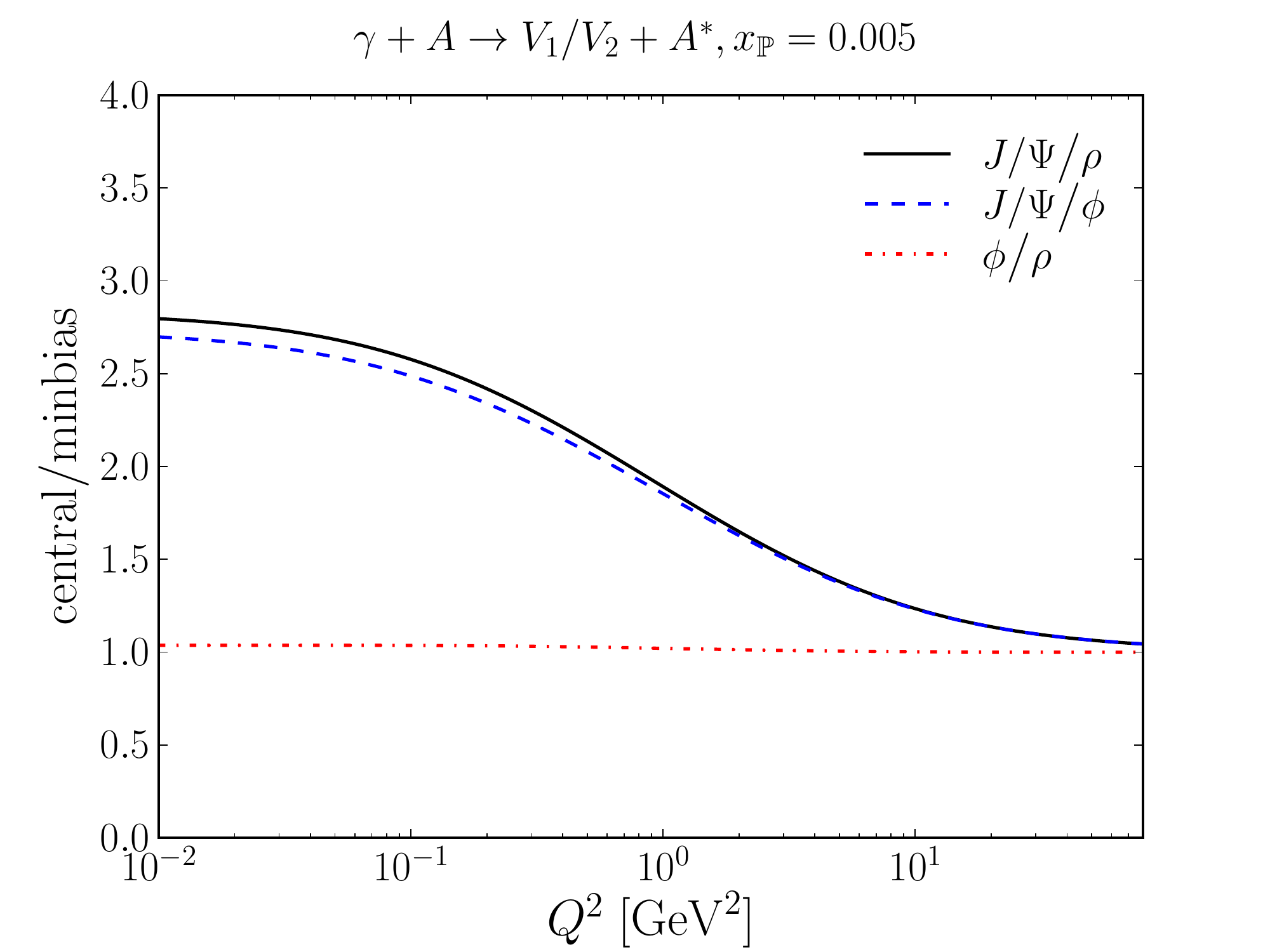} 
				\caption{Ratio of diffractive vector meson production cross sections in central and minimum bias events. Figure from Ref.~\cite{Lappi:2014foa}.}
		\label{fig:centrality_q2}

	\vspace{\baselineskip}%
    \end{minipage}
    
\end{figure}

In addition to measurements of average partonic content of the proton and nuclei, and the event-by-event fluctuations of the density profile, it is also possible to study rare partonic configurations in diffractive processes. One possible way to probe the structure of the nuclei in the region where gluon densities are maximally large (very large saturation scale $Q_s^2$) is to study diffractive scattering as a function of centrality. In incoherent diffractive process, the relatively large $|t|$ ensures that the momentum transfer is localized in the nucleon-size area, and one can define centrality classes.

This idea was developed in Ref.~\cite{Lappi:2014foa}, where the centrality is related to the number of the so called \emph{ballistic} protons. In incoherent diffraction, one nucleon receives a relatively large $p_T$ kick, and it propagates out of the large nucleus. On its way out, it can scatter off the other nucleons. The more central the photon-nucleus interaction is, the nucleon kicks on average more nucleons out of the nucleus. These ``ballistic'' neutrons end up in the zero degree calorimeters and can not be distinguished from ``evaporation'' neutrons that are a result of the nucleus decaying from the excited state long after the scattering process. The ballistic protons, on the other hand, have different transverse momentum than the evaporation protons which follow a thermal spectrum in the rest frame of the nucleus. It was shown in Ref.~\cite{Lappi:2014foa} that the ballistic protons end up in roman pot detectors where their multiplicity is proportional to the centrality.

In the most central events, the saturation scale is significantly larger than on average in the minimum bias events. Thus, one can expect to see a strong suppression in the light vector meson  production cross section compared to $J/\Psi$ production in the events where the proton multiplicity in the Roman pot is maximal. This is demonstrated in Fig.~\ref{fig:centrality_q2}, where the ratio of production cross sections for mesons $V_1$ and $V_2$ in central events (defined as $|b_T|=0$ fm), compared to the minimum bias events (integrated over all $b_T$), is shown. Namely, one calculates
\begin{equation}
\frac{\left. \sigma^{\gamma^* A \to V_1 A^*} \Big/ \sigma^{\gamma^* A \to V_2 A^*}\right|_\text{central}}{\left. \sigma^{\gamma^* A \to V_1 A^*} \Big/ \sigma^{\gamma^* A \to V_2 A^*}\right|_\text{min. bias}}.
\end{equation}
As shown in  Ref.~\cite{Lappi:2014foa}, in case of $J\Psi$-light meson ratio, the double ratio approaches  $Q_{s, \text{central}}^4/Q_{s, \text{min. bias}}^4$ at low $Q^2$, which makes this observable especially sensitive probe of large gluon distributions in the center of the nucleus. 

\section{Summary}

Exclusive and inclusive diffractive processes are powerful tools to probe the small-$x$ structure of the nuclei in the future Electron Ion Collider. The advantages are the especially good sensitivity to gluon densities (to the first approximation, cross section is proportional to the \emph{squared} gluon distribution), and the possibility to access the impact parameter profile of various targets. The possibility to simultaneously study, in addition to average quantities, the event-by-event fluctuations and rare partonic configurations makes diffractive processes even more interesting probes of the small-$x$ dynamics.

\section*{Acknowledgements} 
This work is supported under DOE Contract No. DE-SC0012704. Useful discussions with T. Ullrich and everyone in the BNL EIC Task Force are gratefully acknowledged.

\bibliographystyle{h-physrev4mod2.bst}
 \bibliography{../../refs}

\end{document}